\documentclass[aps,preprint,amsmath,amssymb]{revtex4}
\usepackage{graphicx,color,slashed}
\usepackage{subfigure}
\def\be{\begin{eqnarray}}
\def\ed{\end{eqnarray}}
\def\non{\nonumber}

\begin{document}

\title{  2TeV Higgs boson and  diboson excess at the LHC  }

\author{ \bf Chuan-Hung Chen$^{a}$\footnote{Email:
physchen@mail.ncku.edu.tw} and Takaaki Nomura$^{a}$\footnote{Email: nomura@mail.ncku.edu.tw} }

\affiliation{ $^{a}$Department of Physics, National Cheng-Kung
University, Tainan 70101, Taiwan  }

\date{\today}

\begin{abstract}
Diboson resonance with mass around 2 TeV  in the dijet invariant mass spectrum is reported by  ATLAS and CMS experiments in proton-proton collisions at $\sqrt{s}=8$ TeV. We propose that the candidate of resonance is a heavy neutral Higgs  $H^0$ or charged Higgs $H^\pm$ and use the extended two-Higgs-doublet (THD) to demonstrate the potentiality. We find that the large  Yukawa coupling  to the first generation of quarks can be realized in THD and the required value for producing the right resonance production cross section is of ${\cal O}(0.06-0.2)$.  Besides $WW/ZZ$ channels, we find that if the mass of pseudoscalar $A^0$ satisfies the jet mass tagging condition $|m_j - m_{Z/W}|< 13$ GeV,  the diboson excess could be also caused by $ZA^0$ or $WA^0$ channel. 

\end{abstract}

\maketitle

A  resonance of around 2 TeV in the dijet invariant mass spectrum is recently reported by ATLAS with the data collected at $\sqrt{s}=8$ TeV and 20.3 fb$^{-1}$~\cite{Aad:2015owa}, where each jet is recognized as resulting from a boson decay.  A moderate excess at the same mass region is also found by CMS~\cite{Khachatryan:2014hpa}. Since the tagged jet mass $m_{j}$ is determined by $|m_j -m_V|<13$ GeV, the reconstructed boson could be $W$ or $Z$ in the standard model (SM). The resultant significances at ATLAS in the region around 2 TeV for $WZ$, $WW$ and $ZZ$ channels are $3.4\sigma$, $2.6\sigma$ and $2.9\sigma$, respectively. The associated cross sections $\sigma(p p\to R)BR(R\to VV')$  are in the region of $16-30$ fb, where $R$ is the resonance and $V^{(\prime)}$ is the weak gauge boson $W/Z$.  In order to interpret the diboson excess, the possible candidates are 
a spin-2 Kaluza-Klein mode of the bulk Randall-Sundrum graviton~\cite{Aad:2015owa}, composite spin-1 particle~\cite{Fukano:2015hga, Carmona:2015xaa, Franzosi:2015zra}, spin-1 bosons e.g. $W'$/$Z'$~\cite{Hisano:2015gna,Cheung:2015nha,Dobrescu:2015qna,Alves:2015mua, Gao:2015irw,Thamm:2015csa,Brehmer:2015cia,Cao:2015lia,Cacciapaglia:2015eea,Abe:2015jra,Abe:2015uaa, Dobrescu:2015yba}, and composite spin-0 and/or spin-2 particles~\cite{Chiang:2015lqa,Cacciapaglia:2015nga, Sanz:2015zha}. A possible interpretation by triboson mode is also discussed in Ref.~\cite{Aguilar-Saavedra:2015rna}.

We propose another alternative, where  the resonance of TeV scale is the heavy scalar boson and it  could be a neutral or charged particle. 
In conventional approach, due to the small couplings to the light quarks, the difficulty for a scalar to be the resonance candidate is the low production cross section. We will show how the large couplings of scalar to the first generation of quarks work in the framework of two-Higgs-doublet (THD). The same idea could be applied to more general multi-Higgs  models. Besides the gauge couplings of heavy neutral scalar $H^0 W^+ W^-$ and $H^0 ZZ$, which are similar to the SM Higgs gauge couplings, THD also provides the new interaction $H^0 Z {\cal A}^0$ with ${\cal A}^0$ being the pseudoscalar boson. Basically, if  the mass of ${\cal A}^0$  satisfies the jet mass tagging at ATLAS, i.e. $m_{{\cal A}^0} \sim m_{W/Z}$, we see that $Z {\cal A}^0$ channel could also make a contribution to the fully hadronic final states.  
However,   a light ${\cal A}^0$ in THD is excluded due to the width of $H^0$ being over the narrow resonance requirement (NRR)  $\Gamma_R \lesssim {\cal O}(100)$ GeV~\cite{Aad:2015owa}.  The tension of problem could be easily relaxed by extending the Higgs sector. For instance, a light complex scalar singlet mixes with Higgs doublets.  Due to the mixing effect, the width of $H^0$ decaying into a light $A^0$ and $Z$ then could match the ATLAS limit. We note that since the gluon sub-jets provide unbalanced sub-jet momenta and have a higher number of charged-particle tracks ($n_{\rm trk}$), 
 ATLAS also applies the sub-jet analysis and imposes a cut on the $n_{\rm trk}$ to reduce QCD backgrounds. Hence, when the $A^0$ decays into quark jets,  the cut efficiency of $A^0$ should be same as that of $Z$ . 

As known, the custodial symmetry is preserved in multi-Higgs-doublet models and $\rho = m^2_W/m^2_Z \cos^2\theta_W =1$ is guaranteed at the tree level. Due to the custodial symmetry, the interaction $H^{\pm} W^{\mp} Z$ is forbidden. However, the coupling $H^{\pm} W^{\mp} {\cal A}^0$ in THD is allowed. If $m_{{\cal A}^0} \sim m_{W/Z}$, the events from $W {\cal A}^0$ channel will be similar to those from $WZ$.  Similar to the case of decay $H^0\to Z {\cal A}^0$, the decay width of $W{\cal A}^0$ channel in THD is over NRR of ATLAS. Therefore, we need to extend the Higgs sector to get a light pseudoscalar.   Hence, in this work we are going to explore the potentiality of $H^0$ or $H^{\pm}$ as the 2 TeV resonance and investigate the influence of a  pseudoscalar with mass of ${\cal O}(100)$ GeV.

At first, we demonstrate how the couplings of a scalar to the first generation of quarks can be large in the THD model. We start writing the Yukawa sector of quarks  as \cite{Ahn:2010zza,Chen:2011wp}
 \be
-{\cal L}_Y &=& \bar Q_L Y^U_1 U_R \tilde H_1 + \bar Q_L Y^{U}_2 U_R
\tilde H_2  \non
\\
&+& \bar Q_L Y^D_1 D_R H_1 + \bar Q_L Y^{D}_2 D_R H_2+ h.c.
\label{eq:Yu}
 \ed
with $\tilde H_k=i\tau_2 H^*_k$. By recombining $H_1$ and $H_2$, the new doublets are expressed by
 \be
 h &=&\sin\beta H_1 + \cos\beta H_2 = \left(
            \begin{array}{c}
              G^+ \\
              (v+h^0 +i G^0)/\sqrt{2} \\
            \end{array}
          \right) \,, \non \\
 H &=& \cos\beta H_{1} - \sin\beta H_2=\left(
            \begin{array}{c}
              H^+ \\
              (H^0 +i {\cal A}^0)/\sqrt{2} \\
            \end{array}
          \right)\,, \label{eq:hH}
 \ed
where $\sin\beta=v_1/v$, $\cos\beta=v_2/v$, $v=\sqrt{v^2_1 + v^2_2}$, $\langle H \rangle=0$ and $\langle h \rangle=v/\sqrt{2}$. As a result, Eq.~(\ref{eq:Yu}) can be rewritten by
 \be
 -{\cal L}_{Y} &=& \bar Q_L \bar Y^U_1 U_R \tilde h + \bar Q_L \bar Y^U_2 U_R \tilde H \non\\
 &+& \bar Q_L \bar Y^D_2 D_R h - \bar Q_L \bar Y^D_1 D_R H + h.c.
 \label{eq:Yu2}
 \ed
with
 \be
 \bar Y^{U(D)}_{1(2)} &=& \sin\beta Y^{U(D)}_1 + \cos\beta Y^{U(D)}_2 \,, \non\\
 \bar Y^{U}_{2} &=& \cos\beta Y^{U}_1 - \sin\beta
 Y^{U}_2\,,\non\\
 \bar Y^{D}_{1} &=& -\cos\beta Y^{D}_1 + \sin\beta
 Y^{D}_2\,,
 \ed
where $\bar Y^{U(D)}_{1(2)}$ are related to the mass matrices of  quarks while $\bar Y^{U(D)}_{2(1)}$ provide the couplings of new neutral and charged Higgses to the SM particles. Since the couplings to leptons are irrelevant issue, we do not further discuss the leptonic couplings.   In terms of Eqs.~(\ref{eq:hH}) and (\ref{eq:Yu2}), the physical mass matrix for quarks is given by
 \be
 m^{\rm dia}_F &=& \frac{v}{\sqrt{2}} V^F_L \bar Y^F_{\alpha}
 V^{F\dagger}_{R}\label{eq:mF}
%
 \ed
where $\alpha=1(2)$ while $F=U(D)$ and $V^F_{L,R}$ are the unitary matrices for diagonalizing the quark mass matrix. Clearly, if $\bar Y^{F}_{1(2)}$ and $\bar Y^{F}_{2(1)}$  cannot be diagonalized simultaneously, the flavor changing neutral currents (FCNCs) at tree level will occur and the associated effects are related to the doublet $H$.

It is found that  if $\bar Y^F_{1(2)}$ and $\bar Y^F_{2(1)}$ exist some nontrivial relation, not only could FCNCs  be avoided but also new scalars have unusual couplings to quarks~\cite{Ahn:2010zza,Chen:2011wp}. To see how this happens, we set 
 \be
  I_{12} &=&\left(
         \begin{array}{ccc}
           0 & a & 0 \\
           b & 0 & 0 \\
           0 & 0 & c \\
         \end{array}
       \right)\,,  \
 I_{31}= \left(
         \begin{array}{ccc}
           0 & 0 & a \\
           0 & b & 0 \\
           c & 0 & 0 \\
         \end{array}
       \right) \label{eq:Imatrx}
 \ed
where $a$, $b$ and $c$ are arbitrary complex numbers. Multiplying the mass matrix of  Eq.~(\ref{eq:mF}) by $I_{ij}$ following  $(M^{\rm dia}_F)_{ij}\equiv I_{ij} m^{\rm dia}_{F}I^{T}_{ij}$, we get  
\begin{widetext}
 \be
  (M^{\rm dia}_F)_{12} &=& \left(
         \begin{array}{ccc}
           a^2 m_{f2} & 0 & 0 \\
           0 & b^2 m_{f1} & 0 \\
           0 & 0 & c^2 m_{f3} \\
         \end{array}\right)\,, \
  (M^{\rm dia}_F)_{31} = \left(
         \begin{array}{ccc}
           a^2 m_{f3} & 0 & 0 \\
           0 & b^2 m_{f2} & 0 \\
           0 & 0 & c^2 m_{f1} \\
         \end{array}\right)\,. \label{eq:mFij}
 \ed
\end{widetext}
 Besides  the diagonal forms are preserved, the diagonal matrices of Eq.~(\ref{eq:mFij}) may not have the same mass hierarchy as shown in Eq. (\ref{eq:mF}). Since  there are many  possible $I_{ij}$,  here we just show two examples. The detailed discussions could be referred to~\cite{Ahn:2010zza,Chen:2011wp}.  As a result, the Yukawa couplings of new Higgs bosons to the first generation of quarks could be of order one in principle.
 
 Hence, the couplings of $H^0$ and $H^\pm$ to quarks in THD could be formulated by
 \be
 {\cal L}_{H} &=& \left( \bar d {\boldsymbol \eta}_D P_R d - \bar u {\boldsymbol \eta}^\dagger_U P_L u\right)\frac{ H^0 + i {\cal A}^0}{\sqrt{2}} \non \\
 &+&  \bar u \left( \boldsymbol {V  \eta}_D P_R +  \boldsymbol {  \eta^\dagger _D V^\dagger}P_L \right) d\, H^+ + h.c.\,, \label{eq:Yu}
 \ed
 where ${\bf V}$ is Cabibbo-Kobayashi-Maskawa (CKM) matrix  and ${\rm dia}\boldsymbol{\eta}_{U}=(y_u, y_c, y_t)$ and ${\rm dia}\boldsymbol{\eta}_{D}=(y_d, y_s, y_b)$ are free parameters. For producing TeV  $H^0$ or $H^\pm$ in proton-proton collisions, we set ${\rm dia}\boldsymbol{\eta}_{U(D)}\approx (y_{u(d)}, 0, 0)$ and $y_{u(d)}$ is of ${\cal O}(0.1)$.  As mentioned earlier, we need to modify the THD model to get a light pseudoscalar. Although it is not our purpose to establish a complete model in this paper, however, the simplest extension is to introduce a light complex scalar $SU(2)_L$  singlet. By the mixture with Higgs doublets, the light pseudoscalar could couple to gauge bosons. Hence, by referring to the structure of gauge interactions in THD, we parametrize the relevant couplings of  scalars as~\cite{Gunion:1989we}
 \be
{\cal L} &\supset&  i g m_W c_X W^+_\mu W^{-\mu} H^0 + i \frac{g m_Z}{2\cos\theta_W}  c_X Z_\mu Z^\mu  H^0 - \frac{g \xi s_X (p_{H^0} + p_A)^\mu }{2\cos\theta_W } Z_\mu A^0 H^0\non \\
&+&  \left[\frac{g \xi (p_{H^+} + p_{A})^\mu}{2} W^+_\mu  A^0 H^-  + h.c. \right] \,, \label{eq:IG}
\ed
where $A^0$ denotes the light pseudoscalar, $g$ is the gauge coupling of $SU(2)_L$, $\theta_W$ is Weinberg angle,  $s_X= \sin(\beta -\alpha)$, $c_X = \cos(\beta - \alpha)$, angle $\alpha$ is the mixing angle of two CP-even scalars in THD, and $\xi$ stands for the mixing effect of ${\cal A}^0$ and  light pseudoscalar. If we take $m_{H^0/ H^\pm} \approx 2$ TeV as an input, the involving new free parameters are $y_{u,d}$, $c_X$, $\xi$ and $m_{A^0}$. We note that although the interactions $H^0 (h^0 h^0, A^0 A^0)$ are allowed, however the decay rates are suppressed by $(v^2/m^2_{H^0}, \xi^4 v^2/m^2_{H^0})$, hereafter we ignore their effects. In addition, the coupling $H^\pm W^\mp h^0$ may cause a large width for $H^\pm$. We find that with $c_X \lesssim {\cal O}(0.1)$ and $m_{H^\pm}=2$ TeV, we get $\Gamma(H^\pm \to W^\pm h^0) \lesssim 20$ GeV. Since $h\to b\bar b$ dominates and ATLAS does not find the excess from the b-jet, we do not further discuss its effect.  

\begin{figure}[hptb] 
\begin{center}
\includegraphics[width=73mm]{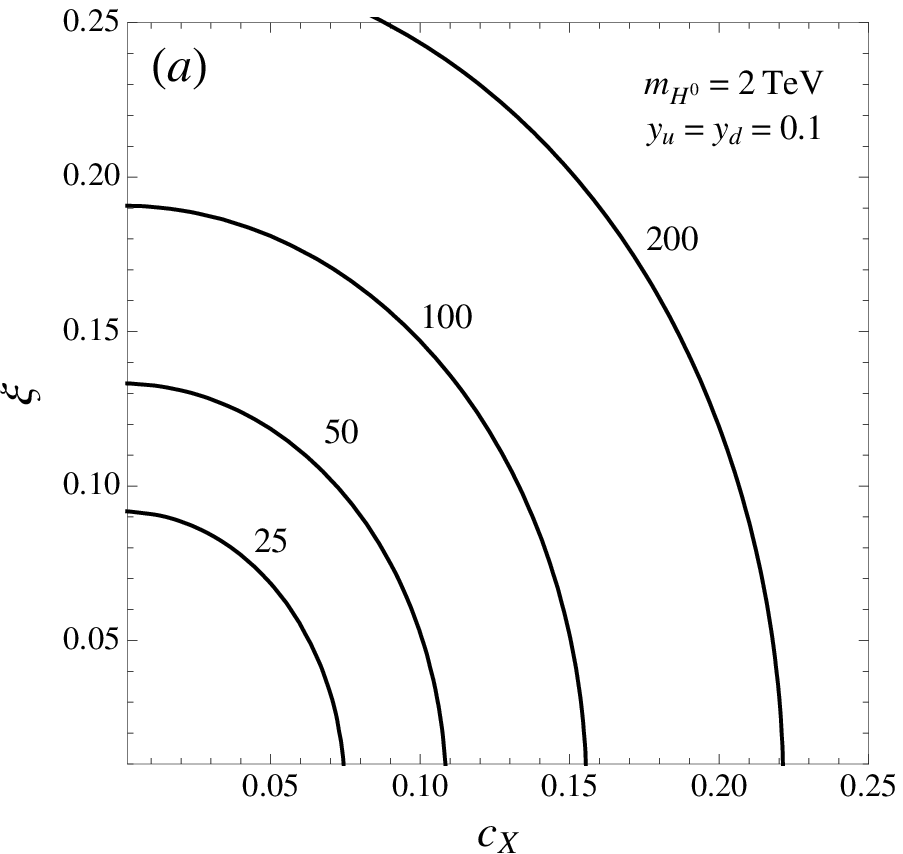} 
\includegraphics[width=70mm]{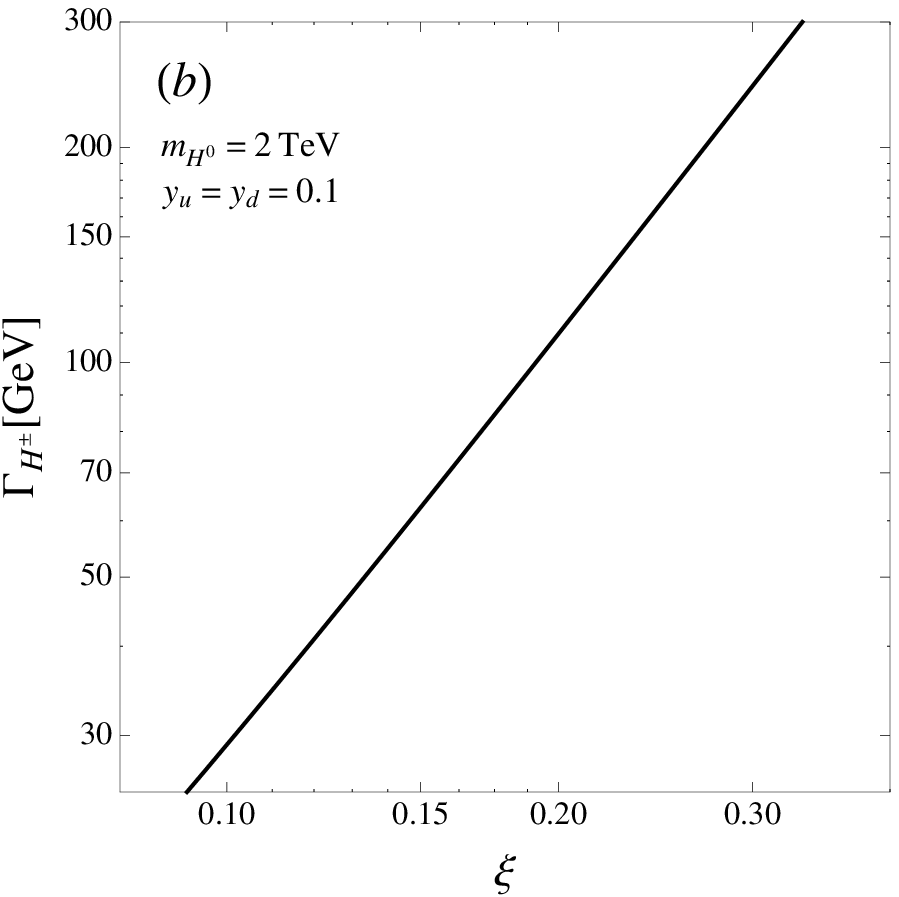} 
\caption{ (a) Contours ( in units of GeV) for $\Gamma_{H^0}$ as a function of $\xi$ and $y_q$. (b) $\Gamma_{H^\pm}$  as a function of $\xi$. In both plots,  we have set $m_{H^0/H^\pm}=2$ TeV and $y_q = y_u =y_d=0.1$.   }
\label{fig:Gamma}
\end{center}
\end{figure}

In terms of the introduced couplings in Eq.~(\ref{eq:Yu}) and the adoption of $y_{u,d}$, the hadronic decay rates of $H^0$ and $H^\pm$  are formulated by
 \be
 \Gamma(H^0 \rightarrow u \bar u (d \bar d)) &=& N_c \frac{y_{u(d)}^2}{16 \pi} m_{H^0}\,, \nonumber \\
\Gamma(H^{+} \rightarrow u \bar d ) &=& N_c \frac{y_{u}^2 + y_{d}^2}{32 \pi} m_{H^\pm} \label{eq:hadronic}
 \ed
 with $N_c=3$ being the number of color. When the scalar mass is fixed to be 2 TeV, the free parameters are only the new Yukawa couplings $y_{u,d}$. By Eq.~(\ref{eq:IG}),  the two-body bosonic decay rates are given by
 \be
\Gamma(H^0 \rightarrow W^+ W^-) &=&  \frac{g^2 m_W^2 c_X^2}{64 \pi m_{H^0}} \frac{m_{H^0}^4 - 4 m_{H^0}^2 m_W^2 + 12 m_W^2}{m_W^4} \sqrt{1- \frac{(2 m_W)^2}{m_{H^0}^2}}\,, \nonumber \\
\Gamma(H^0 \rightarrow Z Z) &=& \frac{1}{2} \frac{g^2 m_Z^2 c_X^2}{64 \pi \cos^2 \theta_W m_{H^0}} \frac{m_{H^0}^4 - 4 m_{H^0}^2 m_Z^2 + 12 m_Z^2}{m_Z^4} \sqrt{1- \frac{(2 m_Z)^2}{m^2_{H^0}} }\,, \nonumber \\
\Gamma(H^0 \rightarrow  Z A^0 ) &=& \frac{g^2 \xi^2 s_X^2 m_{H^0}^3}{64 \pi \cos^2 \theta_W m_{Z}^2}  \lambda^{3/2}\left( \frac{m^2_{A^0}}{m^2_{H^0}}\,, \frac{m^2_{Z}}{m^2_{H^0}}\right)\,, \non \\
%
%
\Gamma(H^\pm \rightarrow  W^\pm A^0) &=& \frac{g^2 \xi^2  m_{H^\pm}^3}{64 \pi  m_{W}^2}  \lambda^{3/2}\left( \frac{m^2_{A^0}}{m^2_{H^\pm}}\,, \frac{m^2_{W}}{m^2_{H^\pm}}\right)\,, \label{eq:bosonic}
 \ed
with $\lambda(a\,, b)= 1 +  a^2 + b^2 -2 a -2 b -2 ab$. The main free parameters are $c_X$, $\xi$ and $m_{A^0}$. Since $m_{H^0/H^\pm} \gg m_{Z, W, A^0}$ in our approach, Eq.~(\ref{eq:bosonic}) could be simplified to be
 \be
 \Gamma(H^0 \rightarrow W^+ W^-) &\approx & \frac{1}{2} \Gamma(H^0 \rightarrow Z Z) \approx  \frac{g^2  c_X^2 m^2_{H^0}}{64 \pi m^2_W} m_{H^0}\,, \non \\
 \Gamma(H^0 \rightarrow  Z A^0) &\approx & \frac{m^3_{H^0}}{m^3_{H^\pm}}   s^2_X \Gamma(H^\pm \rightarrow  W^\pm A^0) \approx  \frac{g^2 \xi^2 s_X^2 m^2_{H^0}}{64 \pi m^2_W} m_{H^0}\,. \label{eq:GammaH}
 \ed
The dependence of $m_{A^0}$ is suppressed in all decay rates. By Eq.~(\ref{eq:GammaH}), we find that $\Gamma(H^0\to WW/ZZ)$ could constrain the $c_X$ while $\Gamma(H^0\to Z A^0)$ could bound $\xi$. It is worth mentioning that  with the limit $m_{A^0} \ll m_{H^\pm}=2$ TeV, the process $H^\pm \to W^\pm A^0$ only depends on $\xi$. It is known that in THD model, when $c_X \to 0$ and $s_X\to 1$, the SM-like Higgs couplings will return to the SM. In this circumstance,  if  $m_{A^0}$ satisfies the jet tagging condition $|m_j - m_{W/Z}|< 13$ GeV, the dijet only can  be generated through  $Z A^0$ or $WA^0$ channel. Using Eqs.~(\ref{eq:Yu}) and (\ref{eq:GammaH}), we present the correlation between total width $\Gamma_{H^0}$  and free parameters in Fig.~\ref{fig:Gamma}(a), where we have adopted $m_{H^0}=2$ TeV and $y_q=y_u=y_d =0.1$. By taking $m_{H^\pm}$ and $y_q=0.1$, the total width $\Gamma_{H^\pm}$ as a function of $\xi$ is given in Fig.~\ref{fig:Gamma}(b). 

\begin{figure}[hptb] 
\begin{center}
\includegraphics[width=70mm]{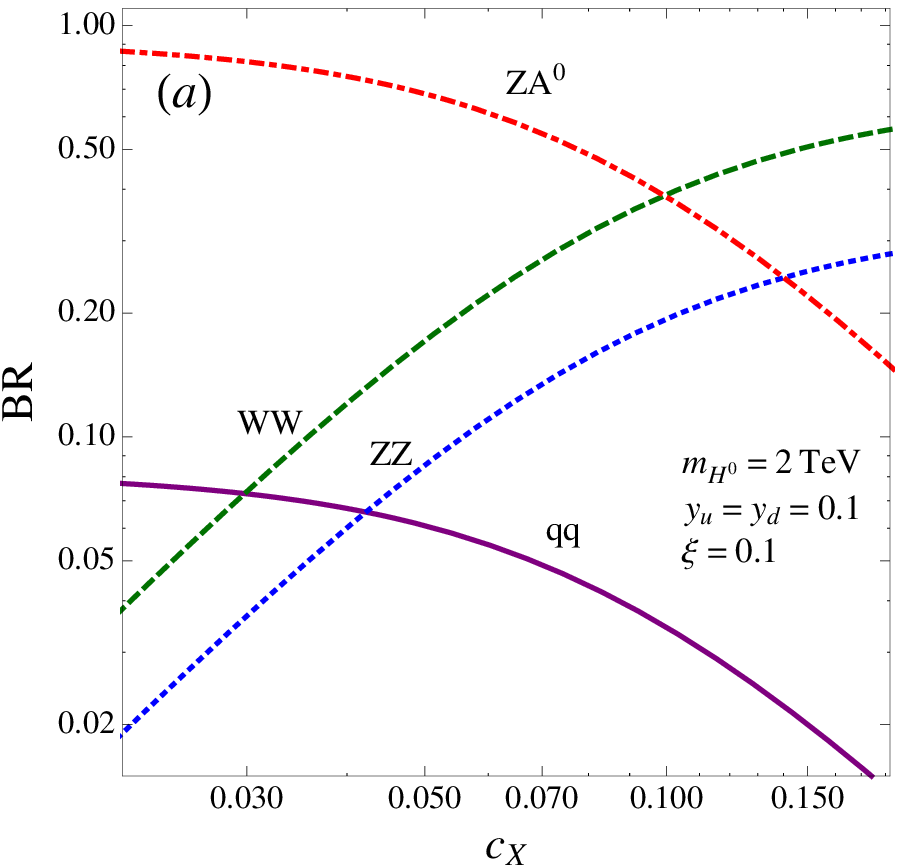} 
\includegraphics[width=70mm]{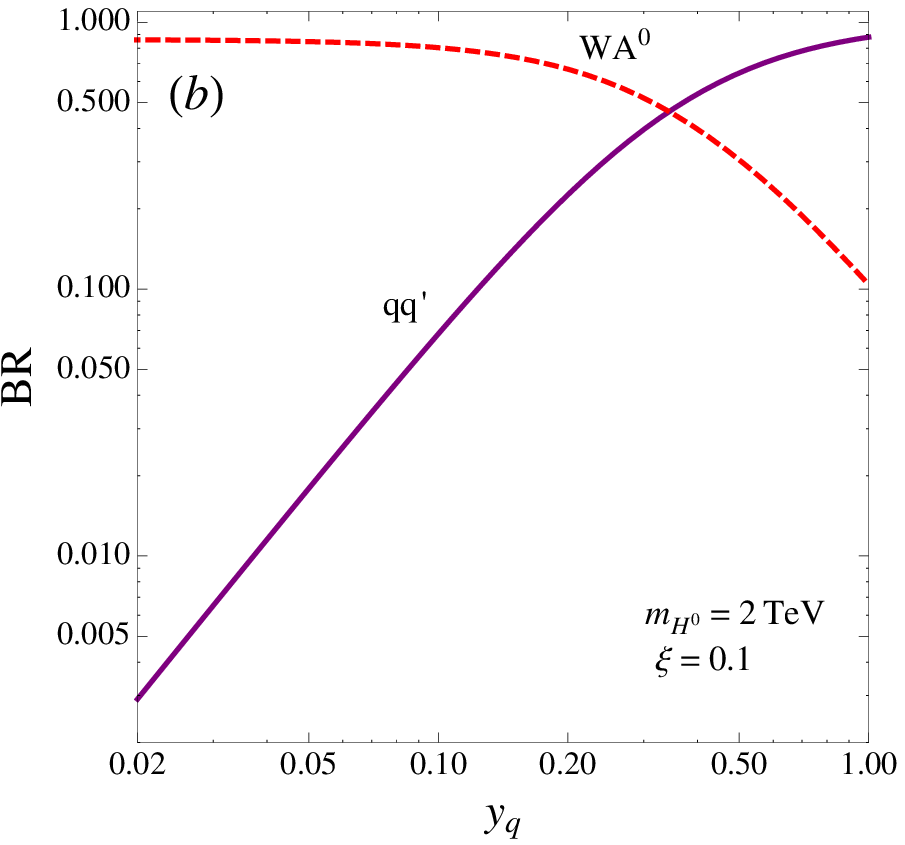} 
\caption{ Branching ratio for (a) $H^0$ and (b) $H^\pm$ decays, where we have set $m_{H^0/H^\pm}=2$ TeV and $\xi=0.1$.   }
\label{fig:BrH}
\end{center}
\end{figure}

It is not clear yet if the resonance is a neutral or charged particle, we separately study the decay branching ratio (BR) and production cross section. According to the decay rates shown in Eqs.~(\ref{eq:hadronic}) and (\ref{eq:GammaH}), we see that $H^0$ can decay into $qq$ with $q=u, d$, $W^+ W^-$, $ZZ$ and $ZA^0$ channels and the  free parameters are $y_q$, $\xi$ and $c_X$.  
We present the BRs for $H^0$ decays as a function of $c_X$ in Fig.~\ref{fig:BrH}(a), where we have used $m_{H^0}=2$ TeV, $y_q =0.1$ and $\xi=0.1$. 
 From the results in Fig.~\ref{fig:Gamma}(a), we see that with $\xi=0.1$,  if $\Gamma_{H^0} < 100$ GeV is required, the value of $c_X$ should be less than 0.14.  In the region $c_X <0.14$, we find that diboson channel dominates the BR.  Accordingly, the results could be consistent with the unseen dijet which is produced from $H^0$ directly. Additionally, we also find that $ZA^0$ mode plays an essential role at $c_X < 0.1$.  

For charged Higgs decays, we only have two channels $W^\pm A^0$ and $qq'$, therefore, the BRs of charged Higgs decays only depend on $y_q$ and $\xi$.  With $m_{H^\pm}=2$ TeV and $\xi=0.1$,  the BRs of $H^\pm$ decays as a function of $y_q $ are displayed in Fig.~\ref{fig:BrH}(b).  In order to avoid the unseen dijet from $H^\pm$ decays,  we should limit $y_q < 0.5$, where $W^\pm A^0$ channel dominates. If we further use $y_q < 0.2$ in which the contribution of $qq'$ mode is small, the BR of $WA^0$ is insensitive to $\xi$. 

\begin{figure}[hptb] 
\begin{center}
\includegraphics[width=70mm]{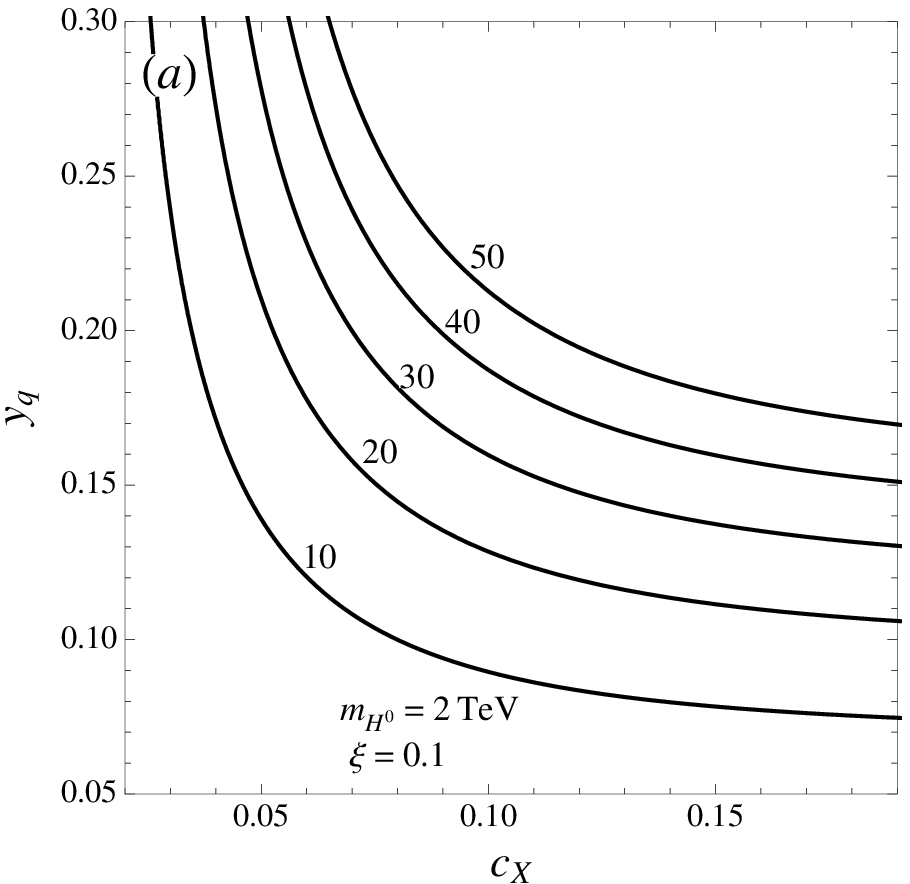} 
\includegraphics[width=70mm]{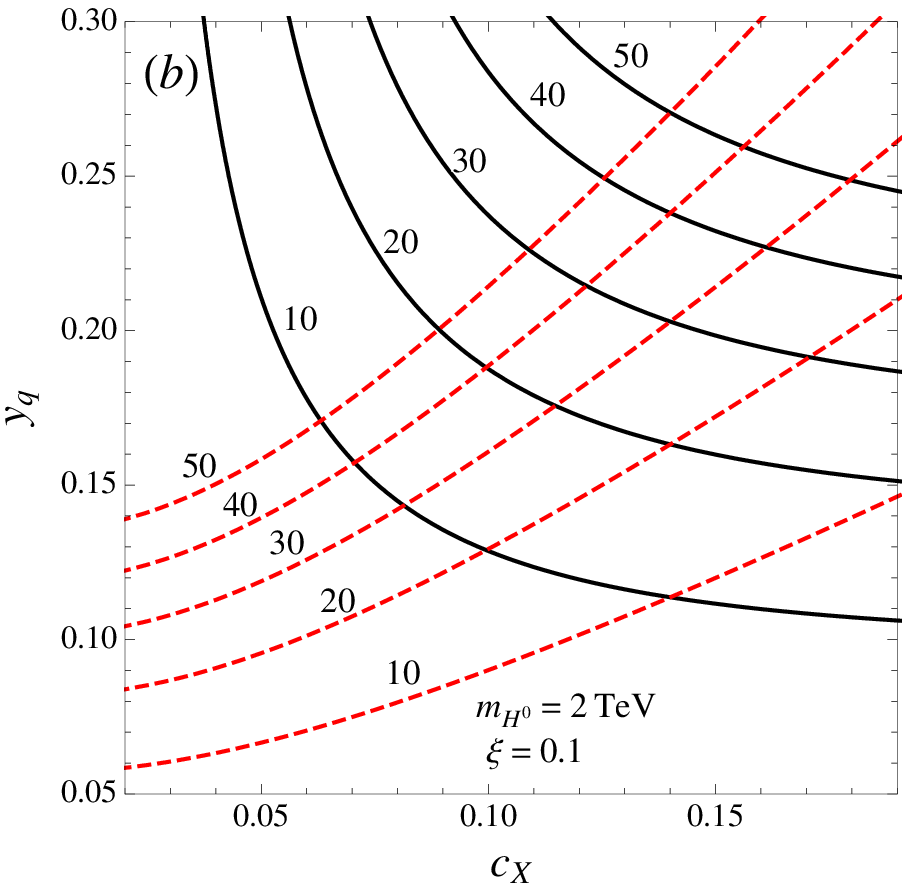} 
\caption{ Contours (in units of fb) for  (a) $\sigma(pp\to H^0)$BR($H^0\to WW$) and  (b) $\sigma(pp\to H^0)$BR($H^0\to Z(Z, A^0)$) as a function of $y_q$ and $c_X$, where the solid and dashed lines in (b) stand for $ZZ$ and $ZA^0$ channels, respectively.    }
\label{fig:CX}
\end{center}
\end{figure}

According to  the number of events observed by ATLAS using 20.3 fb$^{-1}$,  the cross section for $pp\to R \to VV'$ is of order of $16-30$ fb. Based on the results, we investigate if the introduced new interactions can lead to the  same cross section in order of magnitude for $\sigma(pp\to R){\rm BR}(R\to {\rm diboson})$.  For estimating the production cross section $\sigma(pp\to H^0/H^\pm)$, we implement the relevant couplings to CalcHEP~\cite{Belyaev:2012qa} and use it with  {\tt CTEQ6L} PDF~\cite{Nadolsky:2008zw} to calculate the numerical values. By using the cross symmetry, we see that the production of $H^0/H^\pm$ is through quark annihilations and $\sigma(pp\to H^0/H^\pm)$ only depends on $y_q$. Combining the results in Fig.~\ref{fig:BrH}(a), we plot the contours for $\sigma(pp\to R){\rm BR}(R\to {\rm diboson})$ as a function of $y_q$ and $c_X$  in Fig.~\ref{fig:CX}, where  figure (a) is for $WW$ channel and figure (b) is for $ZZ$ (solid) and $ZA^0$ (dashed) channels. 
Moreover,  we find that $y_q\sim {\cal O}(0.15-0.2)$ could match the required cross section. In other words,  the unconventional Yukawa  coupling of scalar to quarks, defined in Eq.~(\ref{eq:Yu}) and adopted ${\rm dia}\boldsymbol{\eta}_{U(D)}\approx (y_q, 0, 0)$, is not necessary to be ${\cal O}(1)$. 

\begin{figure}[hptb] 
\begin{center}
\includegraphics[width=70mm]{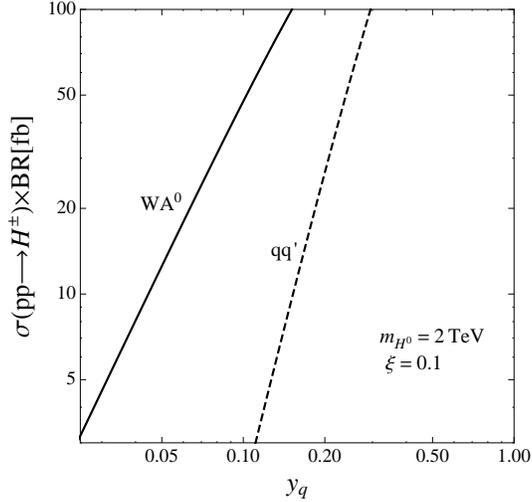} 
\caption{  $\sigma(pp\to H^\pm)$BR($H^\pm\to (W^\pm A^0, qq')$) (in units of fb) as a function of $y_q$.     }
\label{fig:CXHp}
\end{center}
\end{figure}

If the resonance is a charged Higgs, similar to the situation of $H^0$, the production cross section $\sigma(pp\to H^\pm)$ only depends on $y_q$. Although $H^\pm$ can decay into $W^\pm A^0$ and $qq'$, if we focus on  $y_q < 0.1$, the BR of former will approach one while the latter is small and negligible. We show $\sigma(pp\to H^\pm){\rm BR}(H^\pm \to (W^\pm A^0, qq'))$ as a function of $y_q$ in Fig.~\ref{fig:CXHp}. By the plot, it is clear that $y_q\sim {\cal O}(0.06)$ is good enough to interpret  the ATLAS excess. 

Finally, we briefly discuss the constraint from  leptonic  $H^0/H^\pm$ decays. 
According to the measurements of ATLAS~\cite{Aad:2014cka} and CMS~\cite{Khachatryan:2014fba}, the current upper bound on the production cross section of dilepton resonance is known to be ${\cal O}(0.2)$ fb. $\sigma(pp\to H^0)$ of ${\cal O}(10)$ fb is allowed if  the BR of leptonic decay  is less than ${\cal O}(0.01)$, where the required leptonic BR could be achieved when the leptonic Yukawa coupling is $Y_\ell < {\cal O}(0.1)$.
In addition,  the upper bound on the cross section for $pp\to H^\pm \to \ell \nu$  is measured to be ${\cal O}(0.4-0.5)$ fb~\cite{ATLAS:2014wra, Khachatryan:2014tva}.  Like the case for $H^0$, we can escape the constraint if the associated Yukawa coupling satisfies $Y_\ell < {\cal O}(0.1)$. 
In THD models, the Higgs coupling to the lepton sector could be different  from that to the quark sector. If we adopt the type-II THD model, the leptonic Yukawa coupling is $\sim m_\ell \tan\beta /v$. With $\tan\beta\sim 50$ and $v=246$ GeV, one gets $Y_{\mu} \sim 0.02$,  which is much less than ${\cal O}(0.1)$. Hence, the leptonic $H^0/H^\pm$ decays could be consistent with the current upper limits. 

In summary, a diboson excess in dijet invariant mass spectrum  is reported by ATLAS; as a result,  the existence of a resonance with mass around 2 TeV is indicated. We propose the resonance could be the neutral or charged Higgs and employ the extended two Higgs doublets to demonstrate the possibility. We find that the coupling of scalar Higgs to the first generation quarks is not suppressed and the required value to produce a right production cross section for the resonance  is of ${\cal O}(0.06-0.2)$. The involving free parameters are only $y_q$, $\xi$ and $c_X$ when $m_{A^0}\ll m_{H^0/H^\pm}$ and $y_q=y_u=y_d$ are adopted.  Besides the $WW$ and $ZZ$ channels, we find that the channels $ZA^0$ and $WA^0$ in the frame of extended two Higgs doublets are also important. Since the tagged jet mass is only determined by $|m_j - m_{W/Z}|< 13$ GeV, therefore, any new particle  with mass of ${\cal O}(100)$ GeV, e.g. $A^0$ in this approach,  could also contribute to the excess. 
The current limit for light $A^0$ mass in supersymmetric model is  $m_{A^0} > 93.4$ GeV~\cite{Schael:2006cr, PDG2014},  where the LEP data are applied and the related process is $e^+ e^- \to A^0 h$. In our model, the constraint is weaker by following reasons: (1)  $A^0$ production cross section is suppressed by the mixing effect $\xi$; (2) unlike the case   in Ref.~\cite{Schael:2006cr} where  $A^0$ decays into $b \bar b$ or $\tau \bar \tau$, the $A^0$ in the model predominantly decays into light quarks and  the corresponding  background events are larger. 
Therefore,  the  constraint of $m_{A^0}$ in our model should be much weaker and $m_{A^0} \simeq m_Z$ is allowed. 
It will be interesting if a detailed analysis for light $A^0$ of ${\cal O}(100)$ GeV can be searched at the LHC.  \\

\noindent{\bf Acknowledgments}

 This work is supported by the Ministry of Science and Technology of 
R.O.C. under Grant \#: MOST-103-2112-M-006-004-MY3 (CHC) and MOST-103-2811-M-006-030 (TN). 

\end{document}